\documentclass[twocolumn,amsmath,amssymb,prl,superscriptaddress]{revtex4}

\usepackage{graphicx} 
\usepackage{dcolumn}  
\usepackage{bm}       

\begin{document}
\draft

\title{ 
Spin-state transition and spin-polaron physics in cobalt oxide perovskites:\\
{\it ab initio} approach based on quantum chemical methods
}

\author{L. Hozoi}
\affiliation{Max-Planck-Institut f\"{u}r Physik komplexer Systeme,
             N\"{o}thnitzer Str.~38, 01187 Dresden, Germany}

\author{U. Birkenheuer}
\affiliation{Max-Planck-Institut f\"{u}r Physik komplexer Systeme,
             N\"{o}thnitzer Str.~38, 01187 Dresden, Germany}
\affiliation{Forschungszentrum Dresden-Rossendorf, Bautzner Landstr.~128,
             01328 Dresden, Germany}

\author{H. Stoll}
\affiliation{Universit\"{a}t Stuttgart, Pfaffenwaldring 55, 70550 Stuttgart,
             Germany}

\author{P. Fulde}
\affiliation{Max-Planck-Institut f\"{u}r Physik komplexer Systeme,
             N\"{o}thnitzer Str.~38, 01187 Dresden, Germany}

\date\today

\begin{abstract}
A fully {\it ab initio} scheme based on quantum chemical wavefunction methods is used
to investigate the correlated multiorbital electronic structure of a $3d$-metal compound,
LaCoO$_3$.
The strong short-range electron correlations, involving both Co and O orbitals, are 
treated by multireference techniques.
The use of effective parameters like the Hubbard $U$ and interorbital $U'$, $J$ terms
and the problems associated with their explicit calculation are avoided with this 
approach.
We provide new insight into the spin-state transition at about 90 K and the
nature of charge carriers in the doped material.
Our results indicate the formation of a $t_{2g}^4e_g^2$ high-spin state in LaCoO$_3$ 
for $T\!\gtrsim\!90$\,K.
Additionally, we explain the paramagnetic phase in the low-temperature
lightly doped compound through the formation of Zhang-Rice-like O hole states
and ferromagnetic clusters.
\end{abstract}

\pacs{PACS numbers: 71.28+d,71.30+h,72.10-d}
\maketitle

An accurate description of correlated electrons is one of the central problems of
condensed-matter theory.
The commonly used approach for the computation of electronic structures is based on
density-functional theory and various approximations to it, the most applied being the
local density approximation (LDA).
This approach has led to unprecedented progress in computational condensed-matter
physics and surprisingly good results for numerous compounds.
However, it is also known and well documented that the LDA has problems with describing
strongly correlated systems which usually involve $d$ and $f$ electrons. 
For that reason extensions like the LDA+$U$ \cite{LDAplusU} and LDA plus dynamical
mean-field theory (LDA+DMFT) \cite{DMFT_06} were developed.
In the latter approaches, the strong local correlations are described by effective interactions
such as the Hubbard $U$, an interorbital Coulomb repulsion $U'$, and an on-site
exchange coupling $J$. 
Even so, the correlation treatment is limited to on-site effects, despite of the fact
that intersite correlations are known to be sizeable. 
To deal with the latter, the LDA+DMFT scheme has been generalized to a cluster LDA+DMFT
\cite{DMFT_06,DMFT_05}.
But a self-consistent computation of the Coulomb repulsion parameters $U$, $U'$ (and
intersite $V$) is in this context particularly difficult \cite{dmft_U_ferdi04_imada05}.
It is fair to state that the main virtue of LDA, namely its simplicity, has got more
and more lost.

This suggests the use of a different approach, based on the computation of many-body
wavefunctions \cite{fulde_wf_02} by means of quantum chemical techniques \cite{QC_book_00}.
It has the advantage that all approximations are well controlled and that it is not
necessary to introduce effective interactions like $U$, $U'$ or $J$.
Here, we demonstrate this for the case of a $3d$ oxide compound, LaCoO$_3$.
Our analysis is based on a local Hamiltonian approach \cite{IncremScheme_92,graefen_97,shukla_98,CBs_UB_06}
and performed on large fragments including several CoO$_6$ octahedra.
Localized Wannier functions and embedding potentials are obtained for these 
fragments from prior periodic Hartree-Fock (HF) calculations for the infinite crystal
\cite{CBs_UB_06,Crystal_Molpro_interface}.
Since our scheme is parameter-free, variational, and multiconfigurational (or
multireference, see below), it has great predictive power.
It can, for example, predict the basic ingredients required in effective impurity
models like DMFT.
For late transition-metal (TM) oxides, in particular, it is important to anticipate 
whether (and which of) the TM $d$ orbitals are sufficient for constructing a
minimal effective orbital space {\it or} whether both TM $d$ and oxygen $p$
functions must be considered as active.
That such questions can be reliably addressed with quantum chemical methods was
shown before for the layered copper oxides \cite{CuO_qc_07_08} and ladder 
vanadates \cite{NaVO_LH_03}.
However, simple point-charge embeddings were used in previous work for representing
the surroundings of the region where the correlation treatment is carried out.
A newly developed, rigorous embedding technique \cite{CBs_UB_06,Crystal_Molpro_interface}
is applied here for the first time to a $3d$-metal compound.

LaCoO$_3$ has attracted considerable attention due to a number of puzzling phase
transitions induced by changes in temperature \cite{LaCoO_JBG_71,MIT_imada_98}, 
doping \cite{MIT_imada_98,LaCoO_itoh_94}, and/or strain \cite{LaCoO_fuchs_08,LaCoO_rata_08}.
Up to now, most of the experimental work was aimed at understanding the nature of the 
phase transitions in the undoped compound, from a nonmagnetic insulator at low $T$
to a paramagnetic semiconductor above 90 K and to a metal for $T\!>\!500$ K.
The low-$T$ ground-state was assigned to a closed-shell $t_{2g}^6e_g^0$ configuration 
[low-spin (LS), $S\!=\!0$] of the Co ions \cite{LaCoO_JBG_71}.
For $T\!\gtrsim\!90$ K, however, the available experimental results are rather
contradictory. 
While recent x-ray absorption spectroscopy (XAS) \cite{LaCoO_tjeng_06} and inelastic neutron
scattering (INS) \cite{LaCoO_podlesnyak_06} measurements indicate with increasing $T$ 
a gradual transition into a $S\!=\!2$ ($t_{2g}^4e_g^2$) high-spin (HS) configuration of
the Co $3d$ electrons, electron energy-loss spectroscopy data \cite{LaCoO_klie_07}
and the observation of Co-O bond-length alternation \cite{LaCoO_maris_03} suggest 
the formation of a $S\!=\!1$ ($t_{2g}^5e_g^1$) intermediate-spin (IS) state for $T\!>\!90$ K,
susceptible to Jahn-Teller distortions.
On the theoretical side, the transition to a HS state around 90 K is favored by
many-body model Hamiltonian calculations \cite{LaCoO_tjeng_06}.
The IS electron configuration is supported by LDA+$U$ calculations 
\cite{LaCoO_klie_07,LaCoO_korotin_96}.
A wavefunction-based quantum chemical analysis is not available yet for this system.

The first step in our investigation is an {\it ab initio} closed-shell restricted HF
(RHF) calculation for the periodic solid.
The {\sc crystal} code \cite{crystal} is employed for this purpose.
We use all-electron Gaussian-type basis sets of triple-zeta quality \cite{basis_sets}
and a cubic perovskite crystal structure.
The Co-O distance is taken to be $d\!=\!1.92$ \AA , as reported for low $T$ in
ref.~\cite{LaCoO_radaelli_02}.
It is well-known that the HF approximation strongly overestimates the fundamental 
band gap in insulators.
For LaCoO$_3$, the RHF gap is 13.3 eV, where the upper valence bands have O $2p$
character and the lowest conduction bands are related to the Co $3d$-$e_g$ levels.
On-site and intersite correlation effects are here investigated in direct space by
multiconfiguration complete-active-space self-consistent-field (CASSCF) and
multireference configuration-interaction (MRCI) calculations \cite{QC_book_00}.
The orbital basis entering the correlation treatment is provided by the localization
module of the {\sc crystal} program. 
These functions are the Wannier orbitals (WO's) associated with the RHF bands
\cite{zicovich_01}. 
The CASSCF and MRCI calculations are carried out with the {\sc molpro} quantum
chemical package \cite{molpro_2006}.

\begin{figure}[b]
\includegraphics*[angle=0,width=1.00\columnwidth]{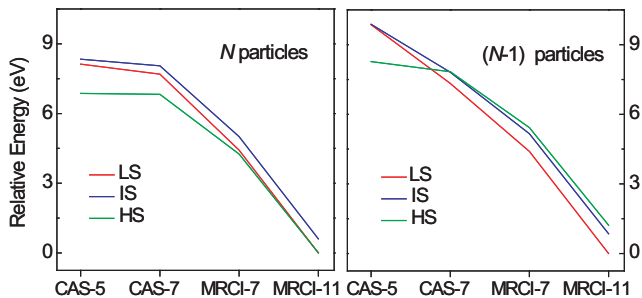}
\caption{
Diagram for the lowest $N$ (left) and ($N\!-\!1$) (right) particle states.
The MRCI treatment is based on a cas-7 reference, see text.
Only the Co $3d$ and O $2p$-$e_g$ electrons are correlated in the
MRCI-7 calculations.
For MRCI-11, the Co $3s$, $3p$ electrons are also included.
The reference is in each graph the energy of the MRCI-11 LS state.
}
\end{figure}

Several sites are included in the correlation treatment.
These sites are partitioned into two groups:
an ``active'' region $\mathcal{C}_A$, which here consists of one CoO$_6$ octahedron, and
a ``buffer'' zone $\mathcal{C}_B$, including the Co and all other O ions of the
six nearest-neighbor (NN) octahedra (i.e., 6 Co and 30 O ions) plus the La NN's of
the ``active'' octahedron (8 La sites) \cite{footnote_1}.
The role of the buffer zone $\mathcal{C}_B$ is to ensure an accurate representation 
of the tails of the WO's centered in the active region $\mathcal{C}_A$.
While the occupied orbitals in the buffer zone are frozen, orbitals centered at sites 
in the $\mathcal{C}_A$ region (and their tails in $\mathcal{C}_B$) are allowed to relax 
and polarize in the quantum chemical calculation.
We denote $\mathcal{C}_A\!+\!\mathcal{C}_B$ as $\mathcal{C}$.
The surroundings of this group of ions (i.e., the crystalline environment) are modelled
as an effective one-electron potential. 
This is obtained from the periodic RHF calculation and incorporated in the one-electron
Hamiltonian in the CASSCF/MRCI treatment via the {\sc crystal-molpro} interface
program \cite{CBs_UB_06,Crystal_Molpro_interface}. 
A ground-state closed-shell self-consistent-field calculation for $\mathcal{C}$ leads 
to changes of only few meV ($\approx\!3$ meV) in the total energy, which shows that our
embedding potential was properly constructed and the projection of the original crystal
WO's centered at sites of the $\mathcal{C}_A$ region onto the set of basis functions
associated with $\mathcal{C}\!=\!\mathcal{C}_A\!+\!\mathcal{C}_B$ \cite{refs_DD} 
leads to insignificant changes of the longer-range ``tails''.

The low-temperature ($T\!\ll\!90$ K) ground-state configuration and the nature of the
lowest $N$-particle excited states are first studied at the CASSCF level.
In CASSCF, a number of active electrons is allowed to be distributed in all possible
ways over a given number of active orbitals.
Both the orbitals and the coefficients in the multiconfiguration expansion are
optimized \cite{QC_book_00}.
In the case of a single Co$^{3+}$ ion, a minimal active space is constructed with
six electrons and five $3d$ orbitals, which we denote as cas-5.
Intriguingly, for such a minimal active space, CASSCF predicts a HS $t_{2g}^4e_g^2$
($^5T_{2g}$) ground-state configuration. 
The LS state, $t_{2g}^6e_g^0$ ($^1\!A_{g}$), is 1.26 eV higher in energy.
The IS state, $t_{2g}^5e_g^1$ ($^3T_{1g}$), is 0.21 eV above the LS state, see Fig.~1.
Clearly, the minimal orbital space description is insufficient for this system.

A first guess is that the active space must be enlarged with O $2p$ orbitals.
Indeed, the RHF band structure, with the O $2p$ bands located above the Co $t_{2g}$
levels, see above, suggests that O $2p$ to Co $3d$ charge-transfer (CT) effects
are important. 
The two $\sigma$-like $e_g$ combinations of $2p$ orbitals at the NN O sites have 
the largest overlap with the Co $3d$ functions.
When these two combinations of $e_g$ symmetry are added to the active space, which
we denote as the cas-7 orbital space, the CASSCF HS-LS splitting is reduced by 0.39
eV, to 0.87 eV.
This is related to the strong interaction between the ``non-CT'' $t_{2g}^6e_g^0$ and
the O $2p$-$e_g$ to Co $3d$-$e_g$ CT configurations in the LS CASSCF wavefunction.
However, the HS state is still the lowest, see Fig.~1.
Other types of charge fluctuations are taken into account by MRCI calculations.
In a first step, we account for all single and double excitations from the Co $3d$
and O $2p$-$e_g$ orbitals. 
The reference is the cas-7 wavefunction. 
The HS-LS splitting is now reduced by an additional 0.69 eV,
with the HS state remaining the ground-state.
A switch in the energy order of these two states is only obtained when accounting
for correlation effects (single and double excitations in the MRCI treatment)
which involve the semi-core Co $3s$, $3p$ electrons.
These correlations are mainly related to the coupling between the O $2p$-$e_g$
to Co $3d$-$e_g$ CT, on-site Co $3s,3p\!\rightarrow\!3d,4s,4p$, and Co
$3d\!\rightarrow\!4s,4p$ excitations.
For such MRCI wavefunctions \cite{footnote_2}, the LS state is found indeed to be the
lowest and the first $N$-particle excited state is the HS state, with an excitation
energy of 6 meV ($\approx\!70$ K, not visible in Fig.~1). 

We thus find that the HS state is lower in energy than the IS arrangement.
For an undistorted lattice, this energy difference is 0.60 eV. 
Additionaly, the quantum chemical calculations provide unique, detailed information
about the importance of different types of correlation effects. 
The good agreement between the LS-HS splitting and the temperature where the magnetic
phase transition occurs (70 vs. 90-100 K) is remarkable.
One should, however, realize that inclusion of other excitation processes and larger
basis sets would modify quantitatively this splitting, although we expect that the order
of the two states would not change \cite{footnote_extra_correl}.

We now proceed with the analysis of the lowest electron-removal and electron-addition
states.
It was argued that the largest interaction with the $2p\!\rightarrow\!3d$ CT
configurations and the strongest stabilization effects occur for the $t_{2g}^4e_g^1$
IS hole state \cite{LaCoO_potze_95}.
Therefore, the lowest ($N\!-\!1$) states are often associated with an IS,
$t_{2g}^4e_g^1$ ($^4T_{1g}$), configuration of the Co ions.
Our calculations show that the lowest-energy hole states are related to 
the LS, $t_{2g}^5e_g^0$ ($^2T_{2g}$), configuration.
Even by CASSCF, with a cas-7 active space, the LS $t_{2g}^5e_g^0$ hole state is about
half eV lower than the IS ($N\!-\!1$) state.
MRCI calculations which include the Co $3s$, $3p$ electrons yield an 
even a larger splitting of 0.85 eV, see Fig.~1.
The implication is that the strongest CT effects occur for reference configurations
with completely empty $3d$-$e_g$ levels, as also found in the $N$-electron
system.

The lowest hole states in LaCoO$_3$ have strong O $2p$ character and resemble 
the lowest ($N\!-\!1$) state in cuprates \cite{ZR_88}.
In cuprates, the dominant contribution to the ionized state is a
superposition of two configurations, $|t_{2g}^{6}d_{z^2}^{2}b^{2}a^{0}\rangle$
and $|t_{2g}^{6}d_{z^2}^{2}b^{0}a^{2}\rangle$, where $b$ and $a$ are bonding and 
antibonding combinations of the O $2p_{\sigma}$ and Cu $3d$ orbitals of ($x^2\!-\!y^2$)
symmetry on one CuO$_4$ plaquette.
In the CASSCF wavefunction \cite{CuO_qc_07_08}, the Cu $3d$ and O $2p_{\sigma}$ 
$(x^2\!-\!y^2)$ orbitals contribute with nearly equal weight to $b$ and $a$.
The added hole has therefore predominant O $2p$ character \cite{CuO_qc_07_08}.
For cubic perovskites, the two $3d$-$e_g$ components are degenerate.
The dominant contributions to the first ionized state in LaCoO$_3$ imply then
three configurations: 
$|t_{2g}^{5} b_{x^2-y^2}^{2} b_{z^2}^{2} a_{x^2-y^2}^{0} a_{z^2}^{0}\rangle$,
$|t_{2g}^{5} b_{x^2-y^2}^{0} b_{z^2}^{2} a_{x^2-y^2}^{2} a_{z^2}^{0}\rangle$, and
$|t_{2g}^{5} b_{x^2-y^2}^{2} b_{z^2}^{0} a_{x^2-y^2}^{0} a_{z^2}^{2}\rangle$. 
The bonding and antibonding $e_g$ orbital combinations are again strong mixtures of
TM $3d$ and O $2p$ functions (see Fig.~2), such that the added hole has large weight
at the O sites.
However, in contrast to the hole state in cuprates, all six ligands are involved,
although in the presence of lattice distortions \cite{LaCoO_maris_03} the $e_g$ levels
split and the hole does not have the same weight at each of the NN ligands.
A study of the coupling between the charge and lattice degrees of freedom in LaCoO$_3$
is left, however, for future work. 

The strong O $2p$ hole character of the ionized state is expected to induce ferromagnetic
(FM) correlations between the Co $t^5$-like ion and its six Co NN's,
like for the $2p$ hole in cuprates \cite{footnote_ZR} or for the O $2p_y^1$ configuration
in NaV$_2$O$_5$ \cite{NaVO_LH_03}.
The formation of FM spin-polaron clusters was indeed inferred from measurements of the
magnetization in lightly doped samples, see, e.g., ref.~\cite{MIT_imada_98}. 
For no or small distortions, the configuration of each of the six Co NN's is HS $t_{2g}^4e_g^2$
and the total spin of the spin cluster is $S\!=\!1/2\!+\!6\!\times\!2\!=\!25/2$.
If the coupling to the lattice is strong, an IS configuration $t_{2g}^5e_g^1$ at the
NN Co sites \cite{LaCoO_maris_03} can be energetically more favorable and the
total spin is $S\!=\!1/2\!+\!6\!\times\!1\!=\!13/2$. 
Obviously, the motion of the added hole is strongly renormalized by the NN FM
correlations.
In cuprates, for example, the NN ``bare'' hopping is reduced by a factor of 4 by short-range
spin interactions \cite{CuO_qc_07_08}.
The study of such correlations require very involved calculations for LaCoO$_3$ because
a number of five $d$ orbitals is needed at each TM site.
A computational approach like that used in copper oxides, where all TM neighbors of the
two octahedra directly involved in the hopping process are included in CASSCF
\cite{CuO_qc_07_08}, is at present not feasible for LaCoO$_3$.
The possibility of using an incremental scheme like that proposed in ref.~\cite{IncremScheme_92}
will be investigated in future work.

\begin{figure}[b]
\includegraphics*[angle=0,width=0.78\columnwidth]{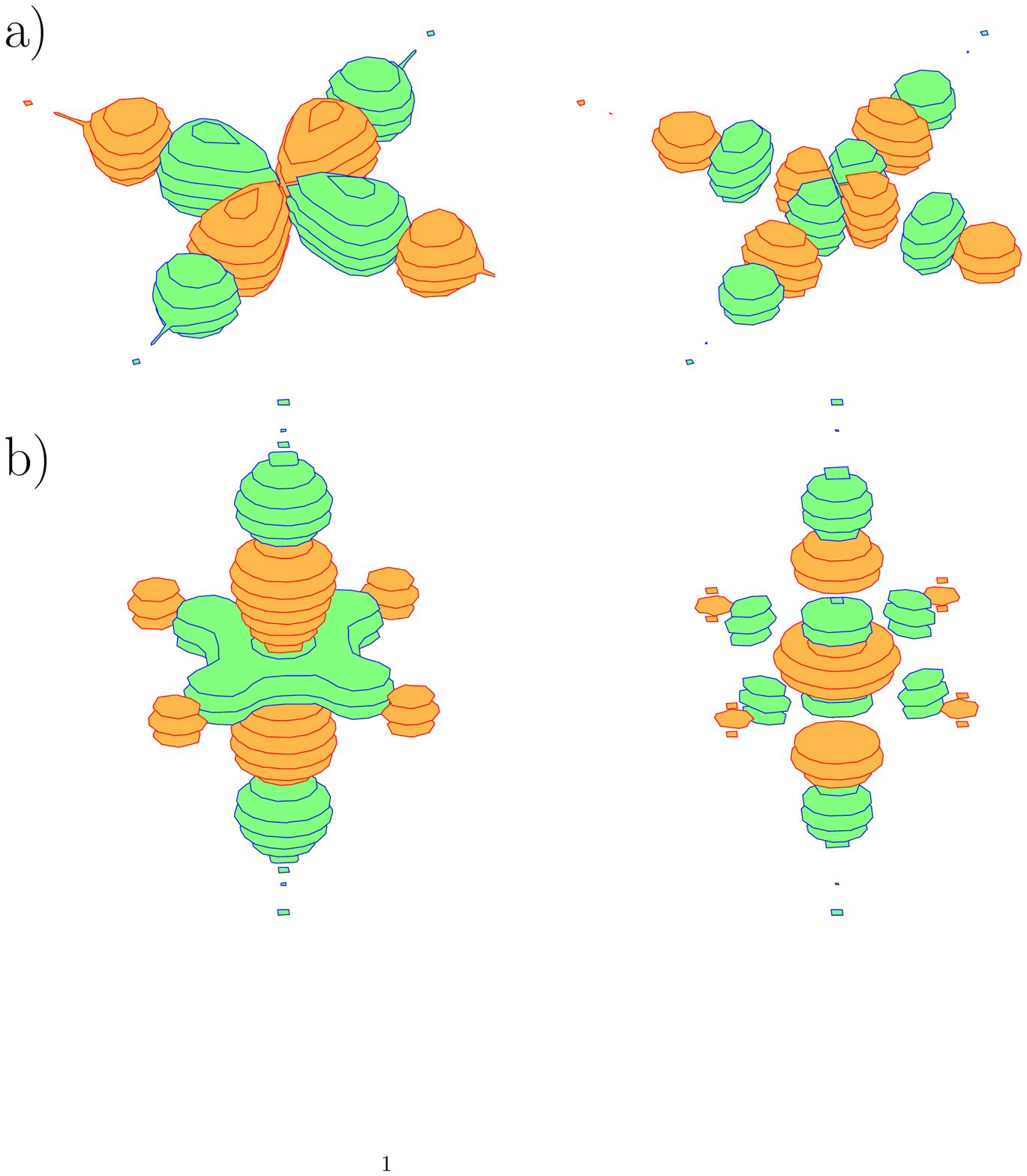}
\caption{Bonding and antibonding $p$-$d$ orbitals for the LS hole state of (a)
$(x^2\!-\!y^2)$ and (b) $z^2$ symmetry.
The $(x^2\!-\!y^2)$ hybrids [see (a)] resemble the ZR $p$-$d$ orbitals in cuprates. 
}
\end{figure}

The lowest $(N\!+\!1)$ states are related to the HS $t_{2g}^5e_g^2$ ($^4T_{1g}$) 
configuration.
MRCI calculations which include in addition to the orbitals used in the cas-7 reference 
the Co $3s$, $3p$ predict a splitting of 1.00 eV between the 
$t_{2g}^5e_g^2$ ($^4T_{1g}$) and $t_{2g}^6e_g^1$ ($^2E_{g}$) states.

The matrix elements (ME's) associated with local electron-removal
and electron-addition configurations like those described above can be used to construct
``correlated'' band structures within a quasiparticle picture \cite{refs_DD}.
The $N\rightarrow (N\!\mp\!1)$ excitation energies correspond to the diagonal (``on-site'')
ME's in the $\mathbf{k}$-dependent secular problem while overlap and Hamiltonian integrals
involving holes or added electrons at different lattice positions $\mathbf{R}$ (i.e.,
$\langle\Psi_{\mathbf{0}n \sigma}^{N\mp 1} |     \Psi_{\mathbf{R}n^\prime\sigma}^{N\mp 1} \rangle$
and
$\langle\Psi_{\mathbf{0}n \sigma}^{N\mp 1} | H | \Psi_{\mathbf{R}n^\prime\sigma}^{N\mp 1} \rangle$
ME's, where $n$ is a band index and $\sigma$ denotes the spin) enter as off-diagonal 
(intersite) terms.
As discussed above, accurate results for the intersite ME's cannot be obtained at this
moment in LaCoO$_3$.
Nevertheless, we can extract from our data estimates for the ionization potential [lowest
$N\!\rightarrow\!(N\!-\!1)$ excitation energy] and electron affinity [lowest $(N\!+\!1)$
state].
In our MRCI calculations, these values are $\mathrm{IP_0}\!=\!4.47$ and
$\mathrm{EA_0}\!=\!-4.24$ eV and incorporate the most important short-range correlations.
Corrections due to long-range polarization effects can be estimated by applying the 
approximation of a dielectric continuum \cite{refs_DD}.
This gives $\mathrm{IP}=\mathrm{IP_0}-C/R$ and $\mathrm{EA}=\mathrm{EA_0}+C/R$,
where $C\!=\!e^2\frac{\epsilon_0 -1}{2\epsilon_0}$, $\epsilon_0\!\approx\!23$ \cite{LaCoO_diel_96}
is the static dielectric constant, and $R$ defines a sphere around the extra
hole or electron beyond which the dielectric response reaches its asymptotic value
$\epsilon_0$.
$R$ is taken here as the average between the NN Co-O and Co-La distances, 
$R\!=\!2.62$ \AA , which leads to $\mathrm{IP}\!=\!1.84$ and $\mathrm{EA}\!=\!-1.61$.
Given the fact that we neglect the finite widths of the two bands, 
the energy difference $\mathrm{IP}\!-\!\mathrm{EA}\!=\!3.45$ eV represents an 
overestimate of the actual band gap.

To summarize, a recently developed {\it ab initio} scheme is applied here to the study of
many-body effects in LaCoO$_3$. 
The nonmagnetic low-$T$ ground-state involves a subtle interplay between ligand-field 
and exchange interactions plus on-site and CT fluctuations. 
The lowest $N$-particle excited state is predicted to be the HS $t_{2g}^4e_g^2$
state.
In agreement with XAS \cite{LaCoO_tjeng_06} and INS \cite{LaCoO_podlesnyak_06} experiments,
the IS $t_{2g}^5e_g^1$ state is 0.60 eV higher.
Whether this HS-IS splitting is sufficiently low to be compensated by the energy gain
associated with a Jahn-Teller ordered configuration of the $e_g$ orbitals, as suggested 
in ref.~\cite{LaCoO_maris_03}, will be the subject of future work.
We also study the nature of the low-energy electron-removal and electron-addition
states.
The lowest $(N\!-\!1)$ excitation is related to a local $t_{2g}^5$, $S\!=\!1/2$ configuration.
The added hole has large weight at the O sites, due to a strong mixing of the
Co and O $e_g$ orbitals.
A large hole density at the O sites causes FM correlations among the adjacent Co
$3d$ electrons,
which agrees with the observation of ``giant'' FM polarons ($S\!=\!10\!-\!16$) and 
paramagnetic behavior at small doping and low $T$ \cite{MIT_imada_98}.
Our analysis offers new insight into the correlated electronic structure of LaCoO$_3$.
Additionally, the {\it ab initio} results provide valuable information for models aiming 
at a realistic description of the finite-temperature properties of this system.

We thank A. D. Rata, M. S. Laad, A. Shukla, G. Maris, and A. Stoyanova for stimulating
discussions.


\begin{thebibliography}{99}
\bibitem{footnote_1} The ``active'' octahedron is centered at (0,0,0), the six NN
octahedra at ($\pm$$a$,0,0), (0,$\pm$$a$,0), (0,0,$\pm$$a$), and the La NN's are at
($\pm$$a/2$,$\pm$$a/2$,$\pm$$a/2$).
\bibitem{footnote_extra_correl} For example, $2p_{\pi}$($t_{2g}$)--$3d$($t_{2g}$)
excitations would slightly favor the HS state, while the use of larger basis sets
should stabilize the LS state.
\end{thebibliography}
\end{document}